# *Amplitude controlled electromagnetic pulse switching using waveguide junctions for high-speed computing processes*

Ross Glyn MacDonald[1,2] Alex Yakovlev[2] and Victor Pacheco-Peña[1*]

[1]School of Mathematics, Statistics and Physics, Newcastle University, Newcastle Upon Tyne, NE1 7RU, United Kingdom
[2]School of Engineering, Newcastle University, Newcastle Upon Tyne, NE1 7RU, United Kingdom
*email: victor.pacheco-pena@newcastle.ac.uk

**Performing computational tasks with wave-based devices is becoming a groundbreaking paradigm that can open new opportunities for the next generation of efficient analogue/digital computing systems. Decision-making processes for switching and routing of signal is fundamental for computing as it enables the transfer of information from one to many (or single) blocks within a system. Here, we propose a technique for the design of pulse-based switching devices for the computing of fundamental decision-making processes. We encode information from multiple channels as transverse electromagnetic (TEM) pulses of varying amplitudes and polarities propagating through interconnected parallel plate waveguides modelled as simple transmission lines. An in-depth description of our technique is presented showing how switching and routing of information can be engineered by exploiting the linear splitting and superposition of multiple pulses traveling through waveguide junctions. To demonstrate the potential of our technique for computing, we develop two devices: a comparator which can calculate the largest value between two real-valued numbers and a pulse director which exploits the reciprocity of waveguide junctions to create a similar yet different performance of a traditional AND gate (emulating its performance via our analogue linear system). These findings may open new pathways for high-speed electromagnetic pulse-based computing systems.**

**Introduction**

In his 1965 paper[1], Gordon Moore described the relationship between the density of transistors on a silicon chip and the operational speed of that chip. In this seminal work, which is now known as Moore's law, he envisioned that the density of these devices would double every 18 months. In the years since its publication, this prediction has been found to be remarkably accurate[2]. This rapid advancement has been possible by the tremendous scalability of semiconductor devices such as metal-oxide-semiconductor field-effect transistors (MOS-FETs)[3,4]. However, Moore recognised that this fast rate of advancement would someday become difficult to continue as there are limits to how small transistor-based devices can be made[5,6], limits that many academics believe we are now reaching[7,8]. The ability to control the flow of energy via switching techniques can be considered as a fundamental process by which computation takes place. For instance, in classical computing devices, the flow of current is controlled by such semiconductor-based devices (transistors) working as switching elements. However, intrinsic parasitic capacitances are difficult to avoid in such devices. These unwanted reactive elements are charged and discharged during the dynamic switching process performed by transistors, adding further delays to the overall computations which restrict the speed and efficiency of the devices[9].

To overcome this, new paradigms on computing are needed. Different scenarios have been recently proposed such as spintronics[10], biological computing[11], computing with electromagnetic (EM) waves[12,13] and optical solitons[14–18], among others. EM wave-based computing has become a hot research topic worldwide as the information can be transferred at the speed of light in the medium where the wave propagates. In this context, such fully EM wave-based computing systems that do not require charging/discharging processes have the potential to open new avenues for future high-speed computing[6].

In this realm, metamaterials (MTMs) and metasurfaces (MTSs) as their 2D version, have been recently applied to the field of computing using EM waves[19–21]. MTMs are artificial materials that exhibit EM responses not always easy to find in natural media such as negative or near zero permittivity values[22–24]. MTMs have successfully been applied in multiple scenarios such as sensing[25,26], antennas[27,28] and imaging[29,30] demonstrating their ability to arbitrarily control fields and waves not only in space but also in time[22,31–37]. This arbitrary manipulation of fields and waves allowed by MTMs has recently been exploited and applied to analogue computing where useful operations such as differentiation, integration and convolution have been demonstrated using waves instead of electrical signals[38–42]. This has inspired the scientific community to propose further approaches for alternative computing using EM waves with outstanding examples including optical quantum computing[43,44] and resonant plasmonic flow networks[45–48]. As with electrical signals in classical computing circuits, switching EM waves is a key process that can be exploited to emulate optical logic operations. Recent examples in this area include the demonstration of logic gates using plasmonic waveguides[49–53], graphene cylindrical resonators[54] and solitons traveling along long chain of polymers[55,56], to name a few. In this realm, we have also recently shown a method for EM pulse switching based on interconnected waveguides in series and/or parallel configurations[57,58]. We have shown how transverse electromagnetic pulses (TEM) pulses of equal amplitude interacting within such structures can constructively/destructively interfere producing new TEM pulses which will or will not propagate down each of the connected waveguides, a response that depends on the polarities (positive or negative) of the TEM pulses used as excitation signals[57,59,60] (i.e., this is a passive system based on linear interferometry[49–51,54,61]).

Inspired by the importance of waves for future computing systems and the need of switching elements as the basis for complex computing processes, in this work we build upon our recently proposed technique to demonstrate a method for high-speed amplitude-controlled switching of EM pulses in multiple waveguide-based junctions. We want to push the boundaries of EM pulse switching by considering junctions being excited by multiple ports at the same time using TEM pulses with different polarities and amplitudes. The underlying physics of the proposed structures are discussed in detail via the scattering matrix approach[59,62]. We provide an in-depth analytical study for the transmission and reflection of pulses between waveguides being connected in series or in parallel. We exploit this technique to demonstrate how these simple, yet interesting waveguide-based junctions can be used to create amplitude-controlled EM pulse switching devices by considering the interaction of pulses in a three-waveguide system. It is shown how it can be implemented for the design of *comparators* which can switch between two states based on the relationship between two inputs. For instance, we discuss how such three-waveguide configurations can be used to compare two numbers ($\varphi_1$ and $\varphi_2$) being mapped as the amplitude of the TEM pulses excited from two different waveguides. It will be shown how the proposed *comparator* can generate a pulse (traveling towards the third waveguide) with either a negative or a positive polarity when $\varphi_1 < \varphi_2$ and $\varphi_1 > \varphi_2$, respectively. We further unleash the potential of our amplitude-based switching approach for the design of an *EM pulse directing* device with the ability to redirect all the pulses in the system (being excited from all the waveguides at the same time) down to a single waveguide, with zero reflection towards the rest of the waveguides. A full physical description of such *directing* device is presented demonstrating how a "matched condition" for the amplitude and polarities of all input pulses should be fulfilled, a condition that also depends on the number of waveguides present in the system. We also provide examples using *N*-interconnected waveguides, for completeness. All analytical results in this study are corroborated by numerical simulations using the transient solver of the commercial software CST Studio Suite® demonstrating an excellent agreement with the theoretical values. In practice these structures could be excited with known microwaves techniques and circuit methods such as line drivers and buffers on transmission lines and using vector network analyzers[63–65]. Our technique for amplitude-based switching of TEM pulses could be exploited in scenarios where decision-making processes are needed such as in the emulation of Boolean logic via analogue linear systems and signal processing tasks.

## Results

### Analytical formulation of the proposed technique

In our proposed technique, decision making processes are performed based on the interaction between TEM square pulses at waveguide crossings (or junctions). The polarity (+ or −) of the transmitted/reflected pulses seen at the waveguide ports can be manipulated by correctly controlling the amplitude of the TEM pulses excited at the ports. As it will be shown, this will enable us to emulate elementary *If ... Then ... Else* operations which are fundamental decision-making processes in computing[66].

To begin with, let us first consider the interaction of TEM square pulses within a network using interconnected 2D parallel plate waveguides (see Fig. 1). As in [57] we consider two types of junctions by using 2D waveguides connected either in parallel or series, see Fig. 1a,b, respectively. Such interconnected waveguides can be represented as transmission lines (TLs)[59] and their equivalent model is shown at the bottom of each network in Fig. 1a,b. In this realm, known TL techniques[59,62] can be exploited to represent the interaction of TEM square pulses when being excited from different ports inside the network. In our approach we consider TEM square pulses of amplitude $A \in \mathbb{R}$ with its sign representing either a positive (+) or negative (−) polarity. Following [57], the polarity of the pulses is mapped by drawing an arrow parallel to a TEM square pulse starting at the base (zero-value) voltage and then directed towards the non-zero voltage (see Fig. 1c). For parallel junctions, pulses with arrows directed towards the top/bottom metallic plate are defined as pulses with a +/− polarity, respectively. For junctions in series, TEM square pulses with arrows directed clockwise/anticlockwise around the junction are mapped as +/− polarity[57].

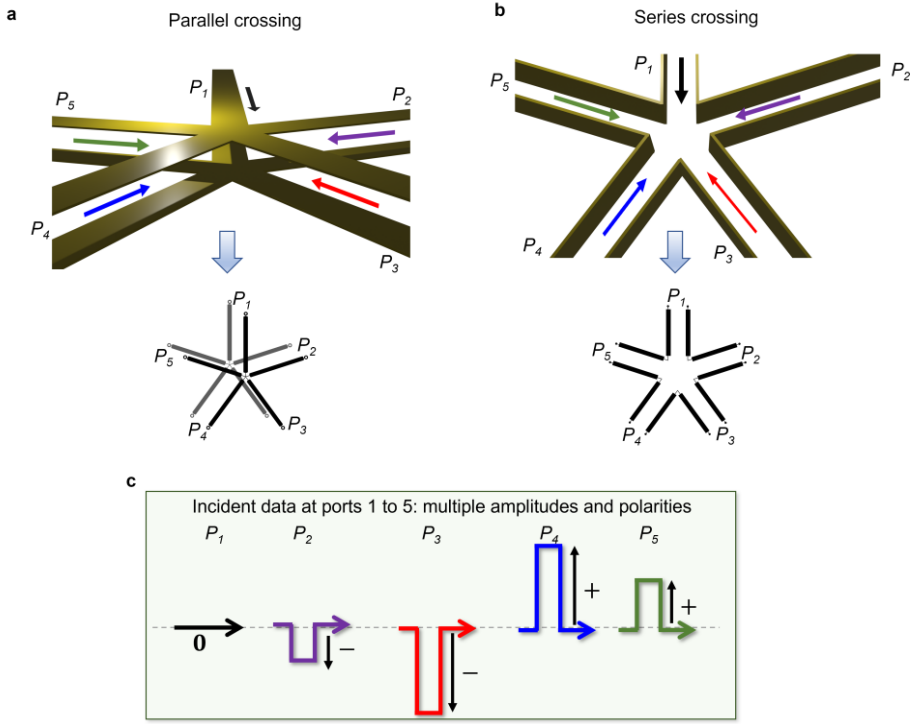

**Fig. 1| Schematic representation of parallel and series waveguide junctions. a,b** 3D render of five-input 2D parallel plate waveguide junctions in parallel and series configurations, respectively. The equivalent transmission line representation of each waveguide junction is shown at the bottom of each panel for completeness. Here the lower/upper set of metallic plates are represented by grey/black cylinders in the transmission line model, respectively, to guide the eye. **c,** examples of different potential input signals which may be excited at each of the waveguide ports labeled as $P_1$-$P_5$.

For $N$-interconnected TLs (as shown in Fig. 1), we can represent the input and output TEM pulses in the network by defining two vectors, namely $\boldsymbol{x} = [x_1, x_2, \ldots, x_N]$ and $\boldsymbol{y} = [y_1, y_2, \ldots, y_N]^T$ respectively, with subscripts 1, 2…$N$ representing each waveguide in the network and $T$ as the transpose operator. In this context, each $x_i$ and $y_i$ element represents the amplitude of the TEM square pulses as described above. The analytical formulation of the $N$-interconnected TLs was recently shown in [57] but we explain its fundamental principles here for completeness noting that this will be further exploited below for decision-making processes using TEM pulses with non-equal amplitudes and polarities. Assuming that all incident TEM square pulses have the same temporal duration and that they arrive at the junction simultaneously (note that we consider that all the wavelengths/TLs are filled with the same materials, air in our case $n_0 = 1$), the vectors $y$ and $x$ can be related by the scattering matrix of the junction, $A$[57].

$$y = Ax^T \tag{1}$$

It is important to highlight that $A$ will generally depend on the geometry and material composition of the waveguides present at the junction. However, when all waveguides are identical, as their impedances are the same, the resulting matrix $A$ can simply be expressed in terms of $N$ (please see the full derivation in the supplementary materials section S1). In this context $A = I - \gamma J$ and $A = -I + \gamma J$ for series and parallel junctions respectively, where $I$ and $J$ are the identity and all-ones matrices of size $N \times N$, respectively, and $\gamma = 2/N$ is the transmission coefficient which depends on the number of waveguides at the junction, as expected. As an example, we can follow this approach to calculate the output vector $y$ of a parallel junction using three interconnected waveguides, resulting in:

$$y = \begin{pmatrix} \frac{-1}{3} & \frac{2}{3} & \frac{2}{3} \\ \frac{2}{3} & \frac{-1}{3} & \frac{2}{3} \\ \frac{2}{3} & \frac{2}{3} & \frac{-1}{3} \end{pmatrix} \begin{pmatrix} x_1 \\ x_2 \\ x_3 \end{pmatrix} = \begin{pmatrix} \frac{-1}{3}x_1 + \frac{2}{3}(x_2 + x_3) \\ \frac{-1}{3}x_2 + \frac{2}{3}(x_1 + x_3) \\ \frac{-1}{3}x_3 + \frac{2}{3}(x_1 + x_2) \end{pmatrix} \tag{2}$$

To better understand the implications of Eq. (2), let us consider an example of such a parallel junction using three waveguides being excited by two single pulses from two different ports simultaneously. This scenario is depicted in Fig. 2 where the amplitudes of the incident pulses are $A$ and $B$ considering a TEM pulse applied from $P_1$ or $P_2$, respectively (here we focus our attention on parallel junctions, the case using series junctions is also included in the supplementary material Fig. S1 for completeness). A schematic representation of this configuration for a time $t$ before and after the pulses have reached the crossing region is shown in Fig. 2a,b, respectively. As observed, the interaction between the two pulses results in six TEM square pulses being *re-emitted* from the junction (three pulses created by each incident TEM square pulse). Pulses which are excited in the same waveguide interact via superposition, resulting into a single output TEM pulse traveling towards each of the ports ($P_1$ to $P_3$). Note that this interaction can be either constructive or destructive depending on the polarities of the *re-emitted* pulses: TEM pulses of opposite polarity will destructively interfere, producing a TEM pulse with lower or even almost zero amplitude. On the other hand, pulses of the same polarity will constructively interact resulting in a single pulse of increased amplitude. This response can be analytically verified by using Eq. (1) with $x = [A, B, 0]$, as follows:

$$y = \begin{pmatrix} \frac{-1}{3} & \frac{2}{3} & \frac{2}{3} \\ \frac{2}{3} & \frac{-1}{3} & \frac{2}{3} \\ \frac{2}{3} & \frac{2}{3} & \frac{-1}{3} \end{pmatrix} \begin{pmatrix} A \\ B \\ 0 \end{pmatrix} = \begin{pmatrix} \frac{-1}{3}A + \frac{2}{3}B \\ \frac{-1}{3}B + \frac{2}{3}A \\ \frac{2}{3}(A + B) \end{pmatrix} \quad (3)$$

As an example, the full range of reachable output pulse amplitudes seen at $P_1$, $P_2$ and $P_3$ when $A$ and $B$ vary in the range $[-1,1]$ volts is presented in Fig. 2c-e, respectively. To better compare these results, we extracted the amplitude of the detected pulses at each output port along the vertical dashed lines from Fig. 2c-e and the results are shown at the bottom of each panel. As observed, the resulting pulses traveling towards $P_1$, $P_2$ and $P_3$ can be positive, negative or even zero depending on the values of $A$ and $B$. In this case, the inflection point between a positive or negative output pulse can be mathematically described by forcing an amplitude of 0V for each of the ports from the right-hand-side of Eq. (3) resulting in $B = 2A$, $B = A/2$ and $B = -A$ for $P_1$, $P_2$ and $P_3$, respectively (to guide the eye these expressions are also plotted as black solid lines in Fig. 2c-e, respectively). Regarding the switching between distinguishable output states, the 0V-lines are of particular importance as we can classify output states based on the polarity of the pulse (output is on one side of the 0V-line or the other, i.e., positive or negative) or based on the existence, or lack thereof, a pulse (output is on the 0V-line or not). Note that the condition $B = -A$ for $P_3$ (representing the bottom waveguide in Fig. 2a,b as the waveguide without an incident TEM pulse) means that the TEM pulses from $P_1$ and $P_2$ should be of opposite polarity but equal magnitude to achieve zero transmission towards $P_3$, an important feature that will be exploited in the following section to demonstrate a comparator as an example of a decision-making process using amplitude controlled switching in $N$-interconnected waveguide junctions.

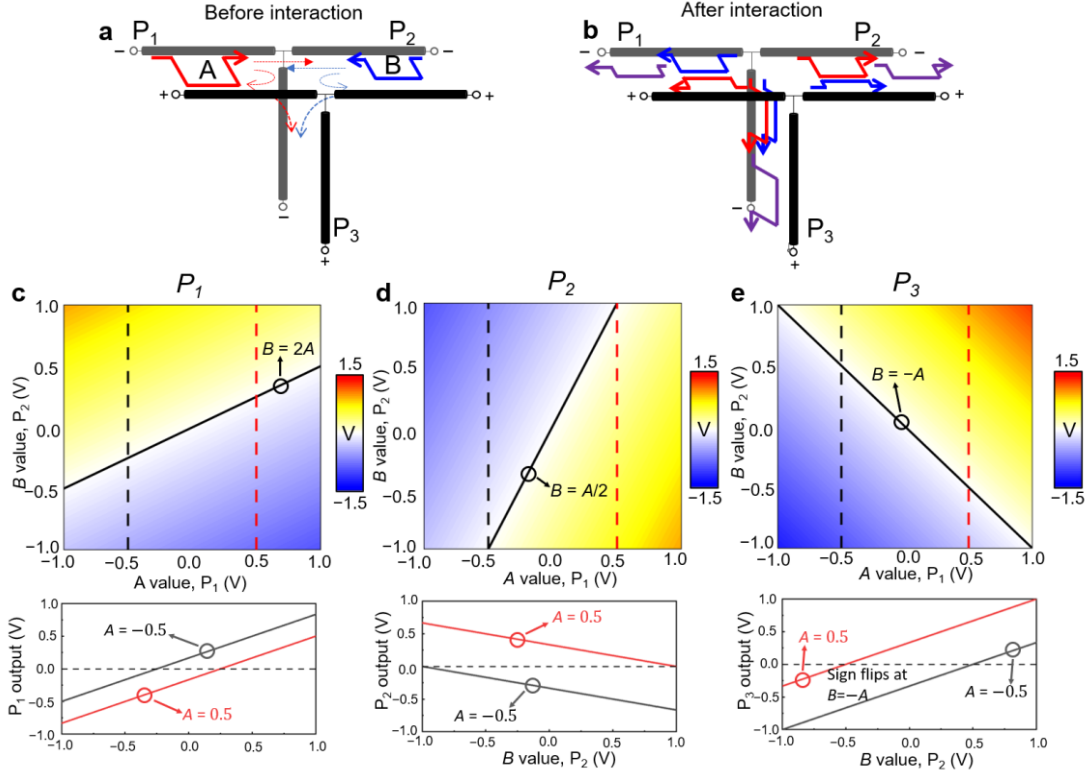

**Fig. 2| Two pulse interaction in a parallel junction formed by three interconnected parallel plate waveguides. a,b** Diagrams representing the interaction of the pulses at times before and after the incident pulses pass the crossing region. Two pulses of arbitrary amplitude are excited from $P_1$ and $P_2$, the splitting of each pulse is represented by arrows in **a**. The individual pulses after splitting are shown in b (red and blue pulses) along with the resulting pulse after superposition (purple pulses). Waveguides are labeled as + and – to help the reader determine the polarity of the pulses at the waveguide port. If the voltage arrow (as shown in Fig. 1c) points towards the waveguide labeled + then the pulse has positive polarity and vice-versa for an arrow pointing towards the waveguide labeled with −. As in Fig. 1 grey/black cylinders in the transmission line models represent the lower/upper set of metallic plates for the waveguides, respectively. **c-e** 2D plots representing the amplitude of the final output pulse observed at $P_1$, $P_2$ and $P_3$, respectively, as a function of the amplitudes $A$ and $B$ of the incident TEM pulses applied from $P_1$ and $P_2$, respectively. The bottom panels in c-e represent the output pulse amplitude seen at $P_1$ - $P_3$ extracted from the black and red dashed lines in **c-e**.

## Comparator

In this section, we exploit three-interconnected waveguides in a parallel junction configuration, as shown in Fig. 2a,b, to design a *comparator* as an example of a decision-making process. The purpose of this structure is to *compare* the values of two real numbers and to decide which one is larger/smaller. In our approach, each number is mapped as the amplitude of the TEM square pulses applied from either $P_1$ or $P_2$. The two numbers we wish to compare are labeled $\varphi_1$ and $\varphi_2$ respectively ($\varphi_1, \varphi_2 \in \mathbb{R}$). The comparator operation is realized by forcing destructive interference between the pulses scattered towards $P_3$ after the incident pulses (one from $P_1$ and one from $P_2$) have passed the crossing region. By observing Eq. (3) such performance can be achieved if the incident pulses from $P_1$ and $P_2$ have opposite polarities, as expected. This is also shown in Fig. 2e where the results of the output pulse at $P_3$ are shown along with the 0V-line (which indicates complete destructive interference) plotted as black solid line. As it can be observed, the 0V-line for $P_3$ has an negative slope (compared to the positive slope for the 0V-lines of $P_1 - P_2$, Fig. 2c,d) indicating that destructive interference occurs when *A* and *B* have opposite polarity, as described above.

To design the *comparator*, we excite a TEM square pulse of amplitude $A = \varphi_1$ and $B = -\varphi_2$ from $P_1$ and $P_2$, respectively. With this configuration, the polarity of the pulse in $P_1$ will thus be the same as the sign of the number the pulse is representing ($\varphi_1$), while the polarity of the pulse in $P_2$ and the sign of the number ($\varphi_2$) are inverted. Following Eq. (3) the amplitude of the output pulse towards $P_3$ can thus be written as:

$$y_3 = \frac{2}{3}(\varphi_1 - \varphi_2) \qquad (4)$$

meaning that the polarity of the output pulse seen at $P_3$ will depend on the values of $\varphi_1$ and $\varphi_2$ such that if $\varphi_1 > \varphi_2$ the TEM pulse will have a positive polarity while it will be of negative polarity if $\varphi_1 < \varphi_2$. Finally, if $\varphi_1 == \varphi_2$ then no pulse will be observed at $P_3$ due to complete destructive interference of the two scattered pulses, as expected[57]. Note that as we exploit the polarity of the output pulse to classify the answer ($\varphi_1 > \varphi_2$ or $\varphi_1 < \varphi_2$), our linear system will only require that the values of $\varphi_1$ and $\varphi_2$ are different enough to produce an output pulse with an amplitude that falls within the dynamic range of a potential receiver/readout. Also, it is important to note that our *pulse comparator* is different than a digital comparator since in our case the inputs and the outputs can have different amplitudes which do not need to be classified as binary 1 or 0, i.e., we are dealing with an analogue system. In this context, it is the polarity (+ or −) not the amplitude of the output pulse what is the result of the decision-making process being carried out by the proposed *pulse comparator*. To verify our *comparator* using waveguide junctions (interconnected TLs) as a decision-making process, full-wave numerical simulations were carried out using the transient solver of the commercial software CST Studio Suite®. TEM pulses with a duration of 0.4ns were excited from $P_1$ and $P_2$. All the waveguides were considered to have the same dimensions (3mm width and a separation between the metal plates of 3mm) and vacuum was used as the filling material ($\varepsilon_r = 1, \mu_r = 1$). The length of the waveguides between each port and the junction is 250mm. With this configuration, the TEM pulses propagate through the waveguides at the speed of light in vacuum (see more details of the numerical setup in the methods section below).

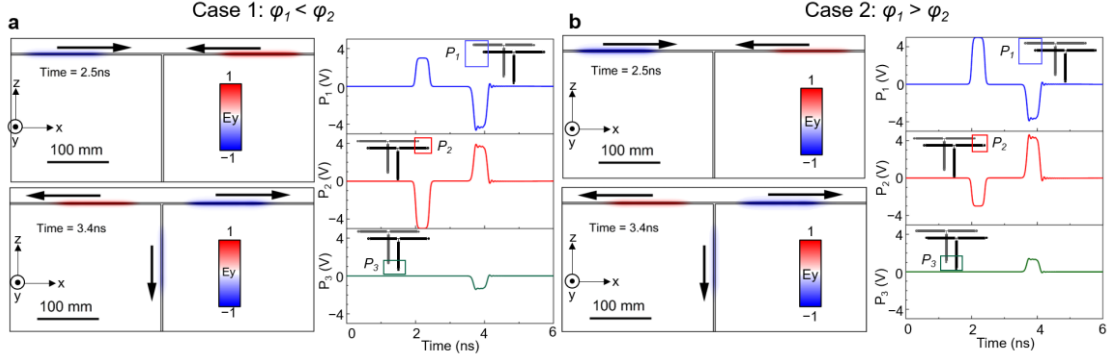

**Fig. 3| Comparator operation, numerical results. a,** case 1, $\varphi_1 < \varphi_2$, considering TEM square pulses at $P_1$ and $P_2$ of are 3V and −5V respectively. **b,** case 2, $\varphi_1 > \varphi_2$, considering TEM square pulses at $P_1$ and $P_2$ of are 5V and −3V respectively. The top-left and bottom-left panels of **a** and **b** represent the snapshot of the out-of-plane electric field distribution at a time $t = 2.5$ns and $t = 3.4$ns corresponding the times before and after the incident TEM square pulses from $P_1$ and $P_2$ have passed the crossing region, respectively). The line plots in the right panel show the voltage as a function of time for each port. All color scales for the out of plane electric field distributions are normalized to 950V/m, corresponding to the maximum electric field obtained from the numerical results for an incident TEM pulse of 3V pulse, which is close to the value expected considering an ideal waveguide with infinite plate size along the transversal $xy$ plane (1000V/m).

The numerical result of the out-of-plane electric field ($E_y$) distribution at different times along with the voltage of the ports as a function of time are shown in Fig. 3. In both Fig. 3a,b a snapshot of the incident pulses at a time instant before ($t = 2.5$ns) and after ($t = 3.4$ns) they interact at the junction is shown in the top and bottom panels, respectively. For completeness, here we consider two possible cases: when $\varphi_1 < \varphi_2$ (case 1, Fig. 3a) and when $\varphi_1 > \varphi_2$ (case 2, Fig. 3b). Without loss of generality, in both cases shown in Fig. 3 the smaller and larger numbers to be compared are chosen to be 3 and 5, respectively (note that our approach also applies for negative values of $\varphi_1$ and $\varphi_2$, see Supplementary information section S2 for an example), resulting in the following incident pulses amplitudes: *(case 1)* 3V at $P_1$, −5V at $P_2$ and *(case 2)* 5V at $P_1$ and −3V at $P_2$. By observing the numerical results from Fig. 3, the amplitude of the pulse traveling towards $P_3$ is −1.3379V and 1.3379V when $\varphi_1 = 3 < \varphi_2 = 5$ (Fig. 3a) and when $\varphi_1 = 5 > \varphi_2 = 3$ (Fig. 3b), respectively. These results are in excellent agreement with the theoretical values predicted by Eq. (4): $-(4/3)$V and $(4/3)$V in $P_3$ for the same two cases in Fig. 3a,b, respectively. These results demonstrate how a comparator between two arbitrary real numbers can be designed by using our approach for TEM square pulse switching via waveguide junctions.

**Pulse director**

In addition to the comparator presented in the previous section as an example of a decision-making process, in this section we will consider an additional device in which all input TEM square pulses are redirected towards a single waveguide with no TEM pulses being reflected to any other waveguide, a structure that we call a *pulse director*. As it will be shown later, such *pulse director* can be generalized to a *N*-waveguide junction. However, we will initially consider the parallel junction with three-waveguides as in the previous section. To find the conditions required for complete pulse redirection towards a single output waveguide, we exploit the reciprocity features of the scattering matrix shown in Eq. (1). In this context, any output pulse vector *y* may be used to reconstruct an input vector *x* by simply using *y* as an input vector[57] (i.e., *x'* = *y*). For example, let us consider the output vector *y* = [0,0,1]V. To find the input vector required to construct this output we can apply the scattering matrix from Eq. (1) to the vector *y* with *x'* = *y,* as follows:

$$x = y' = \begin{pmatrix} \frac{-1}{3} & \frac{2}{3} & \frac{2}{3} \\ \frac{2}{3} & \frac{-1}{3} & \frac{2}{3} \\ \frac{2}{3} & \frac{2}{3} & \frac{-1}{3} \end{pmatrix} \begin{pmatrix} 0 \\ 0 \\ 1 \end{pmatrix} = \begin{pmatrix} \frac{2}{3} \\ \frac{2}{3} \\ \frac{-1}{3} \end{pmatrix} \quad (5)$$

meaning that the input TEM pulses excited in $P_1$ to $P_3$ should have an amplitude of (2/3)V, (2/3)V and −(1/3)V, respectively, to redirect all the pulses towards $P_3$ after they interact at the junction between the waveguides. Clearly this vector *x* = [2/3, 2/3, − 1/3]V is not the only solution to the redirection problem, as the choice of the amplitude of the input pulse for $P_3$ in Eq. (5) was arbitrary. If the calculation in Eq. (5) was repeated for the vector *y* = $\zeta$[0,0,1]V, where $\zeta$ is some arbitrary constant $\in \mathbb{R}$, then *x* = $\zeta$[ 2/3, 2/3, − 1/3 ]V. This means that any input vector that can be written in the form *x* = $\zeta$[2/3, 2/3, − 1/3]V, for any real number $\zeta$, will also result in a pulse redirection effect towards $P_3$. Examples include *x* = [2,2,−1]V and *x* = [1/3, 1/3, − 1/6]V (i.e. $\zeta$ = 3 and $\zeta$ = 1/2 respectively), which results in the output vectors *y* = [0,0,3]V and *y* = [0,0,1/2]V respectively. Here, the *x* vectors that satisfy this condition will be called *matched input vectors*, as the amplitudes and polarities of all input pulses are matched such that only transmission towards $P_3$ is allowed.

As described in the previous sections, after passing the crossing region, the incident TEM pulses applied from each port will create new TEM square pulses traveling in all the waveguides and the final signal that will be received in $P_1$ - $P_3$ will be the result of the destructive/constructive interaction of the pulses inside each waveguide. This is illustrated in Fig. 4a,b where we show a schematic representation of the division of each individual pulse in this system. Here we show the case when the three-waveguide junction is excited with the matched input vector $x$ = [2,2,−1]V as an example. After the pulses pass the junction (Fig. 4b) each incident TEM pulse will generate a pulse traveling in each waveguide towards $P_1$ - $P_3$ with a voltage defined by Eq. (2). However, as the matching condition for vector $x$ is fulfilled, TEM pulses towards $P_1$ and $P_2$ eliminate between each other and only propagation towards $P_3$ is permitted. As shown in Fig. 4b, all the pulses towards $P_3$ have the same polarity and thus are constructively added while in $P_1$ and $P_2$ the pulses completely destroy one another, as expected.

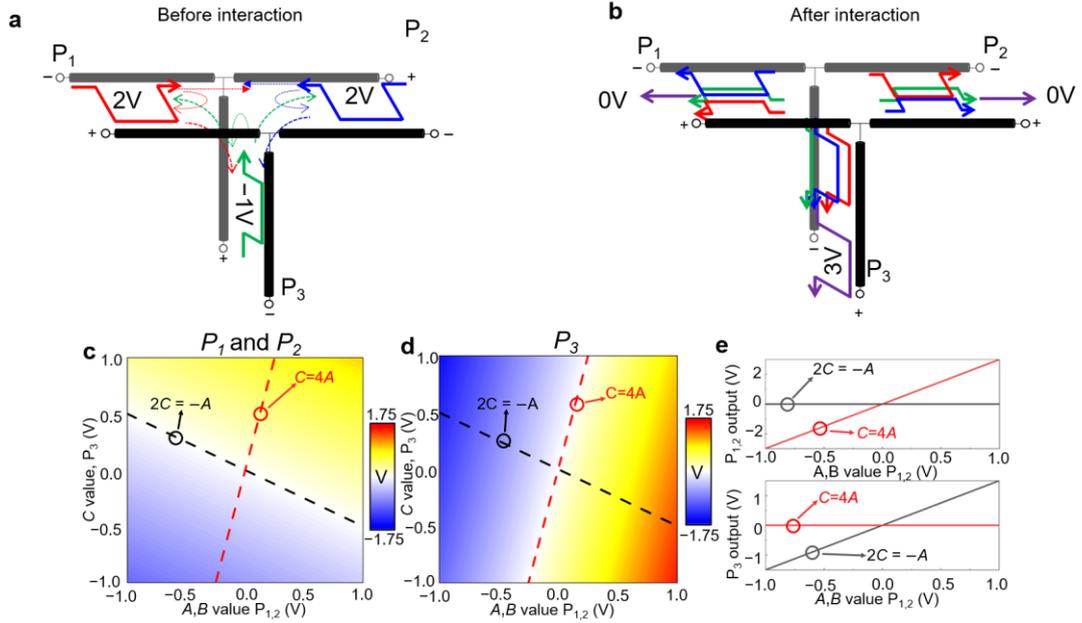

**Fig. 4| Three input pulse director operation.** Pulse diagrams representing the splitting of individual pulses in the matched condition at a time instant a, before and b, after the pulses interact at the junction. Note that, as in Fig. 2, the final pulses traveling towards the ports are the purple pulses. c,d Analytical results of the voltage reachable for the output pulse for $P_1 = P_2$ and $P_3$, respectively, for the case when $A$ = $B$ ∈ [−1,1]V and $C$ ∈ [−1,1]V. As in Fig. 2, the labels + and – have been added to better observe the polarity of the pulses at the waveguide port. As in Fig. 1 the upper/lower waveguides are represented by black/grey transmission lines respectively. **e,** Output pulse amplitudes seen at $P_1 = P_2$ (top) and $P_3$ (bottom) when following the black and red dashed lines shown on the contour plots in c,d.

In Fig. 4c and Fig. 5, we further explore the performance of the structure when the input vector $x$ is chosen such that the *matched* condition is not achieved. Here we consider the input vector $x$ = [$A,B,C$]V where $A$, $B$ and $C$ ∈ ℝ. Given that any *matched* input vector can be written as $\zeta$[2,2,−1] where $\zeta$ is a real constant, we can infer that for $x$ to be matched, $A$ must be equal to $B$. With this in mind, we first explore the range of output pulses seen at the waveguide ports when $A = B$ and $C$ can vary in the range $[-1,1]$V. In this scenario the output pulses seen at $P_1$ and $P_2$ are identical due to the symmetry of the system. With this setup, the amplitude of the output pulses seen at both $P_1$ and $P_2$ are shown in Fig. 4c along with the values seen at $P_3$ in Fig. 4d. As in the previous sections, pulses of +/− polarity are identified within the red/ blue regions, respectively. The 0V-line (the conditions for $A = B$ and $C$ which results in no output pulse seen at $P_1$ and $P_2$) is represented by a black dashed line in the $P_1 = P_2$ plot (Fig. 4c) following the path $2C = -A = -B$ and by a red dashed line in the $P_3$ plot (Fig. 4d) following the path $C = 4A = 4B$. For completeness, the amplitudes of the output pulses seen at $P_1$ and $P_2$ when following $P_3$'s 0V-line and the amplitude at $P_3$ when following the $P_1 = P_2$'s 0V-line are shown in the top and bottom panel of Fig. 4e, respectively. As observed in the top panel of Fig. 4e, when the 0V-line condition of $P_3$ is met (i.e., $C = 4A$) the resulting output pulses at $P_1$ and $P_2$ have an amplitude of $3A$. This can be interpreted as the splitting of the incident pulse from $P_3$ evenly between $P_1$ and $P_2$, as one would expect given that all the waveguides have the same dimensions and filling materials. In this interpretation the incident pulses from $P_1$ and $P_2$ both eliminate the signals towards $P_3$ and enhance the transmitted pulses to $P_1$ and $P_2$ resulting in all the power being evenly divided between $P_1$ and $P_2$. The bottom panel of Fig. 4d shows that when the 0V-line condition for $P_1$ (or $P_2$) is met, the output pulse amplitude seen at $P_3$ is $3A/2$.

Following this setup, let us consider the case where $A$ and $B$ are now free to vary while $C$ is held at $-(1/2)$V. Note that this corresponds to an arbitrary value, as an example; a matched condition can be found for any value of $C$. The range of the possible received signals at each output port is shown in Fig. 5. Here the top panels represent the range of attainable output amplitudes seen in $P_1$, $P_2$ and $P_3$ respectively from left to right. As observed, the symmetry of $P_1$ and $P_2$ is now broken compared to the results shown in Fig. 4, leading to them having separate 0V-lines, namely $B = A/2 + 1/2$, $B = 2A - 1$ and $B = -A + 1/4$ for $P_1$, $P_2$ and $P_3$ respectively. From these results, if one wants to eliminate the pulses traveling towards $P_1$ and $P_2$, the matching conditions for both $P_1$ and $P_2$ must be met simultaneously, as expected. This occurs at the intercept between $P_1$ and $P_2$'s 0V-lines (i.e. $A = 1, B = 1$), as can be seen in the bottom panels of Fig. 5a,b. This is an expected result as the matched vector for $C=-(1/2)$V is $x = [1, 1, -1/2]$V. These results demonstrate that other than the trivial case of $x = [0,0,0]$V, the only input vectors that result in no reflection seen at $P_1$ and $P_2$ are the *matched* input vectors. This performance was verified via numerical simulations considering four different cases for the input vector $x$: i) $x = [2,2,−1]$V, ii) $x = [1,3,−1]$V, iii) $x = [1,2,−1]$V and iv) $x = [2,3,−1]$V. The numerical results for cases i and ii are shown in Fig. 6a,b, respectively. The results for cases iii and iv can be found in the Supplementary materials, for completeness. As observed, the scenario shown in Fig. 6a represents the *matched* condition discussed in Fig. 4a,b while the other scenarios from Fig. 6b and Supplementary materials represent slight deviations from this *matched* configuration. As it is shown, reflections down $P_1$ and $P_2$ are observed in all cases except for the *matched* scenario, corroborating the performance of the proposed *pulse director*. Moreover, note that these numerical results are in excellent agreement with the analytically calculated values found through Eq. (1) where $x = [2,2,−1]$V results in $y = [0,0,3]$V, $x = [1,3,−1]$V to $y = [1, −1,3]$V, $x = [1,2,−1]$V to $y = [1/3, −2/3, 7/3]$V and $x = [2,3,−1]$V to $y = [2/3, −1/3, 11/3]$V. In the numerical results the vectors $y$ are $y = [−0.0049, −0.0049, 3.0042]$, $y = [0.9984, −1.0082, 3.0042]$, $y = [0.3313, −0.6720, 2.3373]$ and $y = [0.6622, −0.3411, 3.6710]$ respectively.

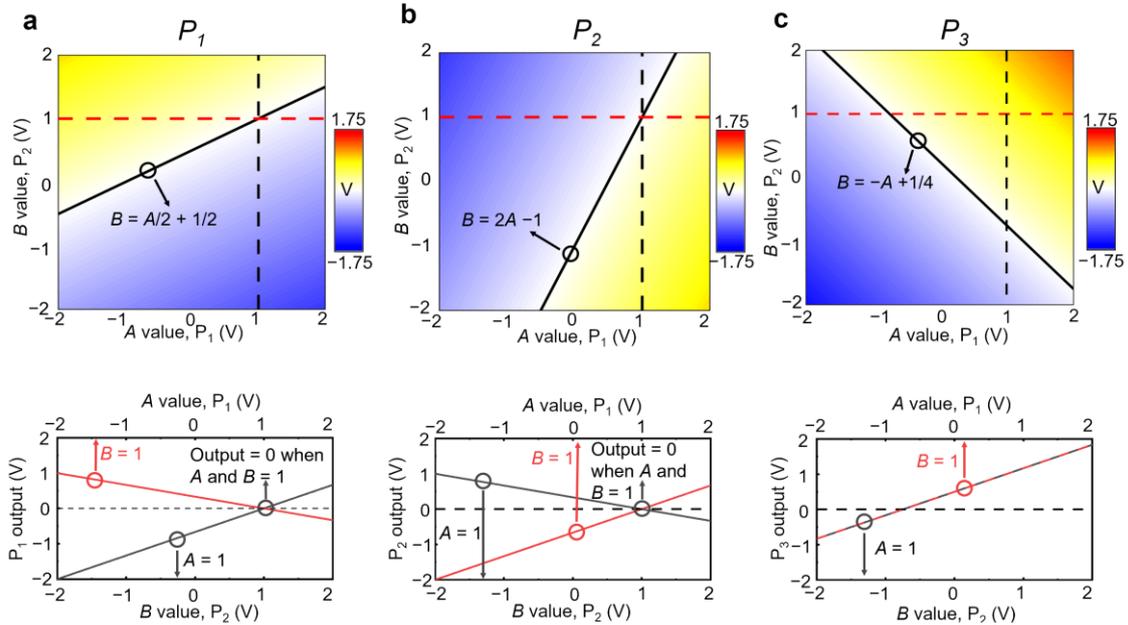

**Fig. 5| Three input pulse director with uneven inputs.** Output pulse amplitudes of the pulse director when $A \in [−2,2]V$, $B \in [−2,2)]V \in$ and $C = −(1/2)V$. a,b,c show the output pulse amplitudes at $P_1$, $P_2$ and $P_3$ respectively. The line plots beneath each panel show the amplitude of pulses seen at $P_1$, $P_2$ and $P_3$ when following the black and red dashed lines shown on the contour plot above.

**Pulse director in *N*-interconnected waveguides**

As we mentioned in the previous section, the proposed *pulse directing* technique can be generalized to a *N*-waveguide junction by following the same reciprocity approach described for the three-waveguide configuration in Fig. 4a,b. Whereas discussions of the comparator and the three-waveguide pulse director in the previous sections have been applied to parallel junctions, in this section we will study the case of *N*-interconnected waveguides using series junctions (see Fig. 1b). The purpose of this is twofold: i) to show how the *pulse directing* technique can be applied to both series and parallel junction setups and ii) to demonstrate the extension of our technique to an arbitrary number of waveguides. As can demonstrated in supplementary material section S1, the scattering matrix for a *N*-waveguide series junction can be written as $A_{series} = -A_{parallel}$. This implies that any vector which is matched for one junction setup, is also matched for the other as the output vector $y$ is simply multiplied by a factor of $-1$.

As described in the previous sections, for an *N*-input waveguide junction any single incident pulse applied from a port will result in the creation of *N* pulses (one per waveguide) after the incident pulse passes the junction. From Eq. (1) it can be seen that for a *N*-waveguide series junction the transmitted (towards the non-incident ports) and reflected (towards the incident port) pulses will have an amplitude $-2/N$ and $(N-2)/N$ multiplied by the incident pulse amplitude respectively[57,62]. For example, considering an incident pulse with an amplitude of 1 V from the $N^{th}$ waveguide, the output pulse vector is $y = [-2/N, -2/N, \ldots, (N-2)/N]$ V. As before, the reciprocity of the system implies that if this vector $y$ is applied as an input vector, then only a single pulse in the $N^{th}$ waveguide would be observed after the incident pulses pass the waveguide junction. In this context, the *matched* vector for a *N*-waveguide junction is thus $x_{matched} = [-2/N, -2/N, \ldots, (N-2)/N]$. As any vector that can be written in the form $x = \zeta x_{matched}$ is also a matched vector, the above *matched* vector can be simplified into the form:

$$x_{matched} = [1, 1, \ldots, (2-N)/2] \qquad (6)$$

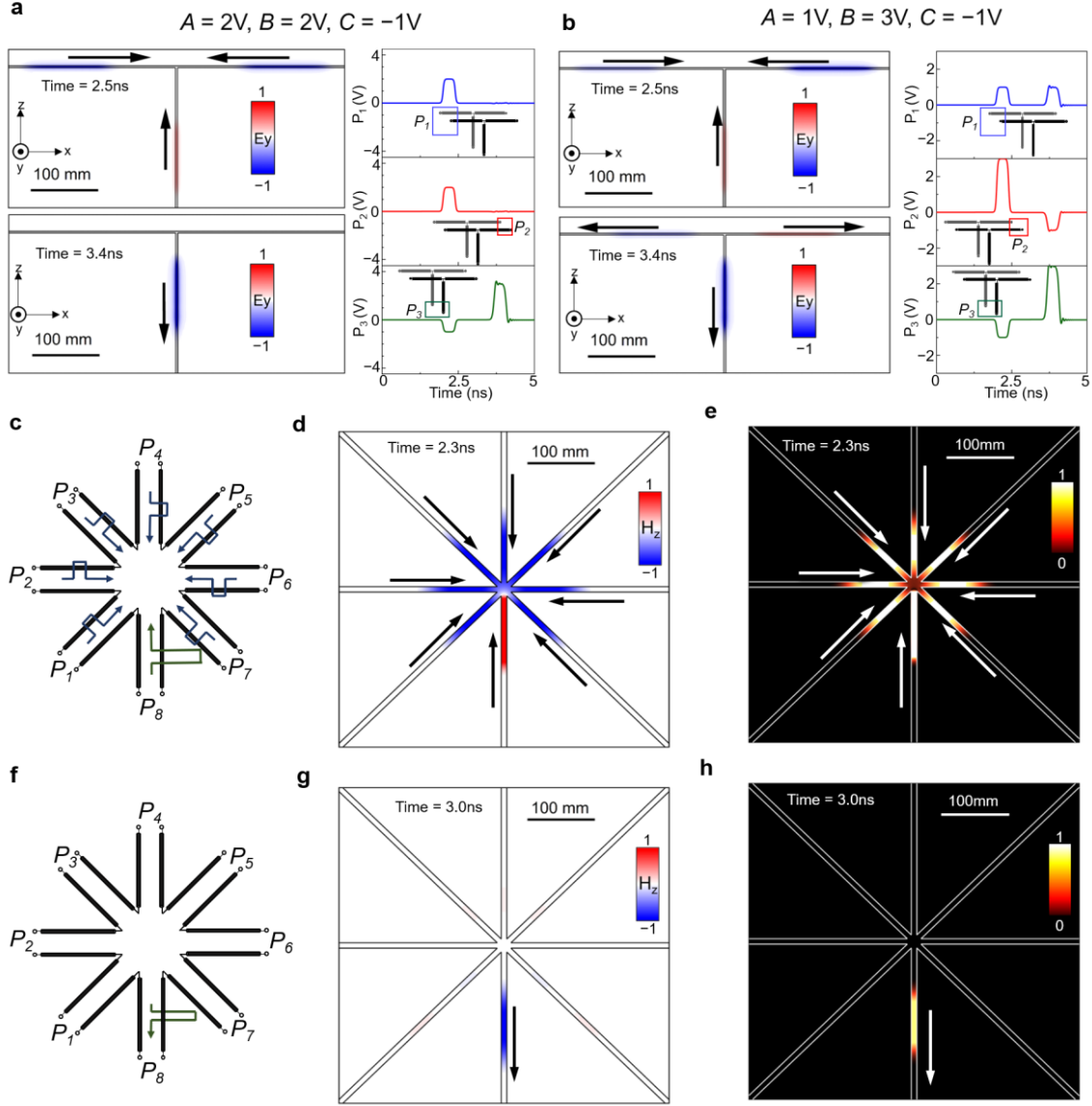

**Fig 6| Pulse director, numerical results. a-b** (left) snapshots $E_y$ field distribution at times before, $t =$ 2.5ns, and after, $t =$ 3.4ns, the incident TEM square pulses from $P_1$ - $P_3$ have passed the crossing region, respectively. All color scales for $E_y$ are normalized to 633V/m, corresponding to the maximum $E_y$ field obtained from the numerical results for an incident TEM pulse of 2V. Note that this value is close to the expected maximum when considering an ideal waveguide with infinite plate size along the transversal $xy$ plane (666.7V/m). The right panels in **a-b** show the voltage as a function of time for all three ports. **a**, $A = 2V$, $B = 2V$, $C=−1V$ and **b**, $A = 1V$, $B = 3V$, $C =−1V$. **c,f** TL representation of an 8-waveguide series junction with ports labeled $P_1$ - $P_8$ for times before and after the pulses have passed the crossing region, respectively. **d-e** snapshot of the normalized $H_z$-field and power distribution, respectively, at a time ($t =$ 2.3ns) before the pulses have reached the junction. **g-h** same cases as **d-e** but at a time after the junction interaction ($t =$ 3.0ns). The bottom waveguide is excited with a −3V TEM pulse and all other are excited with a 1V pulse. This results in a −4V pulse in the bottom waveguide and no other pulses present. The color scales for the panels corresponding to a time before the incident TEM pulses have reached the junction ($t =$ 2.3ns, panels **d** and **e**) are normalized to the maximum value of the out-of-plane $H_z$-field and power distribution values respectively, corresponding to an input pulse of 1V (1A/m and 375W/m$^2$, respectively). The color scales for the results $t = $ 3ns (**g** and **h**) are normalized to 4A/m and 6000 W/m$^2$, respectively, corresponding to the peak values for an incident pulse of −4V.

This performance was corroborated by full wave numerical simulations using the time domain solver of the commercial software COMSOL Multiphysics® (see details about the simulation setup in the methods section below) and the numerical results of an eight-waveguide series junction are shown in Fig. 6d,e,g,h along with the TL representation of this configuration in Fig. 6c,f. Here, we excite ports $P_1$ - $P_7$ with a 1V TEM square pulse and $P_8$ with a −3V pulse. The input $x$ vector of this excitation is then $x$ = [1,1,1,1,1,1,1,−3]V, which corresponds to a matched input vector. Note that $P_8$ is situated at the bottom in all panels in Fig. 6c-h. Fig. 6c-h shows the schematic representation of the equivalent TL (Fig. 6c), normalized out-of-plane $H_z$ field distribution (Fig. 6d) and the normalized power distribution (Fig. 6e) of the incident pulses at a time ($t$ = 2.3ns) before their interaction at the junction. The field distributions and TL sketch at a time ($t$ = 3ns) after the incident pulses have passed the crossing region are shown in Fig. 6f-h. As observed, only one pulse is visible traveling towards $P_8$ with an amplitude of −3.99V, in agreement with the theoretical values using Eq. (1) which predicts an output vector $y$ = [0,0,0,0,0,0,0,−4]V, demonstrating how the proposed decision-making process can be implemented in $N$-interconnected waveguides.

As explained in this work, the *comparator* and *pulse director* devices here studied could be exploited in computing processes as they represent primitive operations. For example, for a three-waveguide junction as those evaluated in the previous sections, the existence, or lack thereof, of reflected pulses observed at $P_1$ and $P_2$, represent two distinct switching states, which in the realm of decision-making processes can be interpreted as *True* or *False* values. As previously discussed in this manuscript, this enables elementary *If...Then...Else* operations to be processed with a high speed (with the velocity of light in the medium filling the waveguides). For example, if a pulse of amplitude −(1/2)V is excited in $P_3$ it is possible to emulate an AND structure between the pulses excited in $P_1$ and $P_2$. In this case the matched vector will be $x$ = [1,1,−1/2]V, which leads to the conditional structure: "*If* (Pulse amplitude in $P_1$) == 1V AND (Pulse amplitude in $P_2$) == 1V *then* return (No reflected pulses), *else* Return (Reflected pulses)". In this case "(No reflected pulses)" would be interpreted as *True* and "(Reflected pulses)" would be interpreted as *False.* The structures presented in this work may be scaled down to allow for more compact devices and shorter pulse durations. In such scenarios dispersion of the metals involved in the design should be considered[67], as we have shown in our previous work[57]. These results demonstrate the potential of controlling the switching process of TEM square pulses in interconnected waveguide junctions. We envision that such scenarios may open new avenues in high-speed computing as decision making processes such as *If…Then…Else* are key primitives in computing systems.

**Conclusion**

In this work we have proposed and demonstrated a computing technique that exploits the splitting of TEM pulses of different amplitudes and polarities (representing the information to be processed) being excited from multiple ports and traveling within multiple waveguides connected either in a parallel or series junction configuration. The fundamental physics of TEM pulse propagation in such junctions has been presented and studied using transmission lines techniques, demonstrating how the interaction of multiple incident pulses are the result of the power division and superposition of the incident pulses within the system. To explore the opportunities of our technique for decision-making processes as fundamental for computing systems, two applications have been presented: a *comparator* and a *pulse director*, showing how they can be designed to decern the larger value between two numbers or to achieve similar performances as traditional logic gates (here emulated via an analogue linear system), respectively. We envision that our technique may find applications in wave/pulse-based processors and could be potentially merged with electronic systems and devices (such as CMOS technologies) opening new directions for future high speed computing systems.

**Methods**

The numerical simulations shown in Figs. 3-6 were performed using the transient solver of the commercial software CST studio suite®. Perfect electric conductor (PEC) was used for the metallic plates having a zero thickness. All the waveguides were designed with parallel plates having a transversal dimension of 3mm (width) and a length of 250mm to the center of the junction. For the three interconnected waveguides in parallel, the distance between the plates was 3mm. Vacuum ($\varepsilon_r = 1, \mu_r = 1$) was used as both the filling material of the waveguides and the background medium. Waveguide ports were used to excite the waveguides with pulses of different polarities and amplitudes having a duration of 0.4ns in their non-zero state. For the parallel plate configurations in Figs. 3-6, top and bottom "open-(add space)" boundary condition was implemented to numerically evaluate the performance of the junctions when they are immersed within an infinite vacuum background. The rest of the boundaries were simulated with an "Open" boundary condition. The numerical simulations for the *N*-waveguide series junction shown in Fig. 7 were carried out using the time-domain solver of the commercial software COMSOL Multiphysics® using the same materials as in the results discussed in Fig. 3-6. In this case, a distance of 10mm between the plates was considered. Scattering boundary conditions were used at the entrance of each waveguide to both excite each of them with an incident square pulse (duration of 0.4ns and fall/rise time of 0.08ns with a second derivative smoothing) and to absorb the pulses traveling towards the waveguides after the incident pulses have passed the crossing region.

## Acknowledgements

V.P-P. and A. Y. would like to thank the support of the Leverhulme Trust under the Leverhulme Trust Research Project Grant scheme (RPG-2020-316). V.P.-P. also acknowledges support from Newcastle University (Newcastle University Research Fellowship). V.P-P. and R.G.M would like to thank the support from the Engineering and Physical Sciences Research Council (EPSRC) under the scheme EPSRC DTP PhD scheme (EP/T517914/1).

## Conflicts of interests

The authors declare no conflicts of interests.

## Author contributions

V.P.-P. and A. Y. conceived the original idea of using interconnected transmission lines for switching of multiple TEM pulses with different amplitudes. R.G.M conducted the numerical simulations and analytical calculations. All the authors were involved in the discussion and interpretation of the results. R.G.M. wrote the first draft of the paper and then V.P.-P and A. Y. commented, edited, and worked on the subsequent drafts of the paper. V.P.-P. supervised the project.